\documentclass[journal]{IEEEtran}

\IEEEoverridecommandlockouts

\usepackage{tikz}
\def\checkmark{\tikz\fill[scale=0.4][color=black!60!green](0,.35) -- (.25,0) -- (1,.7) -- (.25,.12) -- cycle;} 
\usepackage{amsthm}
\usepackage{amsfonts}
\usepackage{amssymb}
\usepackage{mathrsfs}
\usepackage{mathtools}
\usepackage[english]{babel}
\usepackage{float}

\usepackage{amsmath}
\usepackage{xcolor}
\usepackage{url}
\usepackage{multirow}
\usepackage{graphicx}

\DeclareMathOperator*{\argmin}{arg\,min}
 \usepackage{booktabs}
\usepackage{bm}
\usepackage{balance}
\usepackage{microtype}
\usepackage[caption=false,font=footnotesize]{subfig}
\usepackage{cite}
\usepackage{comment}

\usepackage{cuted,tcolorbox}
\definecolor{infocolor}{RGB}{245,245,245}
\usepackage{multicol}
\usepackage{capt-of}

\allowdisplaybreaks

\newtheorem{theorem}{Theorem}[section]
\makeatletter
\newcommand{\settheoremtag}[1]{
  \let\oldthetheorem\thetheorem
  \renewcommand{\thetheorem}{#1}
  \g@addto@macro\endtheorem{
    \addtocounter{theorem}{-1}
    \global\let\thetheorem\oldthetheorem}
  }
\makeatother

\usepackage[switch]{lineno}
\usepackage[named]{algo}
\usepackage[noend]{algpseudocode}
\usepackage{algorithm}

\newcommand{\bsym}{\boldsymbol}

\DeclareMathOperator*{\minimize}{minimize }

\begin{document}

\title{An Invitation to Hypercomplex Phase Retrieval: Theory and Applications}

\author{Roman Jacome~\IEEEmembership{Student Member,~IEEE}, Kumar Vijay Mishra~\IEEEmembership{Senior Member,~IEEE}, Brian M. Sadler~\IEEEmembership{Life Fellow,~IEEE}, Henry Arguello~\IEEEmembership{Senior Member,~IEEE}
\thanks{R. J., and H. A. are with Universidad Industrial de Santander, Bucaramanga, Santander 680002 Colombia, e-mail: \{roman2162474@correo., henarfu@\}uis.edu.co.}
\thanks{K. V. M. and B. M. S. are with the United States DEVCOM Army Research Laboratory, Adelphi, MD 20783 USA, e-mail: kvm@ieee.org, brian.m.sadler6.civ@mail.mil.}
\thanks{This research was sponsored by the Army Research Office/Laboratory under Grant Number W911NF-21-1-0099, and the VIE project entitled ``Dual blind deconvolution for joint radar-communications processing''. K. V. M. acknowledges support from the National Academies of Sciences, Engineering, and Medicine via the Army Research Laboratory Harry Diamond Distinguished Fellowship.}
}

\maketitle
\IEEEpeerreviewmaketitle

\begin{abstract}
Hypercomplex signal processing (HSP) provides state-of-the-art tools to handle multidimensional signals by harnessing intrinsic correlation of the signal dimensions through Clifford algebra. Recently, the hypercomplex representation of the phase retrieval (PR) problem, wherein a complex-valued signal is estimated through its intensity-only projections, has attracted significant interest. The hypercomplex PR (HPR) arises in many optical imaging and computational sensing applications that usually comprise quaternion and octonion-valued signals. Analogous to the traditional PR, measurements in HPR may involve complex, hypercomplex, Fourier, and other sensing matrices. This set of problems opens opportunities for developing novel HSP tools and algorithms. This article provides a synopsis of the emerging areas and applications of HPR with a focus on optical imaging.
\end{abstract}


\section{Introduction}
In many engineering applications, it is useful to represent the signals of interest as hypercomplex numbers, that is, they are elements of some algebras over the field of real numbers. While vector spaces allow only addition and scalar multiplication, algebras permit both addition and multiplication between their elements. In particular, hypercomplex representation enables multidimensional signal and image processing applications by harnessing Clifford algebra to exploit intrinsic correlation within the different signal dimensions. Hypercomplex signal processing (HSP) finds applications in several important problems including optical imaging, array processing, wireless communications, filtering, and neural networks; see, e.g., \cite{miron2023quaternions,buvarp2023quaternion} and references therein. Recently, HSP has been investigated for tackling the long-standing problem of phase retrieval (PR) in high-dimensional settings that usually arise in optical imaging. In this article, we provide an introduction to the fundamental concepts of hypercomplex phase retrieval (HPR) and corresponding sample problems in optical imaging.

\subsection{Historical notes}
Hypercomplex numbers trace back to the nineteenth century with the pioneer works of Sir William Hamilton on quaternion algebra in an 18-instalment seminal paper published in volumes XXV–XXXVI of the Philosophical Magazine between 1844 and 1850 (see, e.g., \cite{hamilton1844ii} and the successive publications in the series). This also included the formulation of biquaternion numbers in the same series. Later, John T. Graves, inspired by Hamilton's work, discovered octonions in 1843. However, Graves published his discovery after Arthur Cayley's well-recognized work on quaternions and their relation with elliptic functions, wherein the latter also described the octonions \cite{cayley1863jacobi}. Leonard E. Dickson unified these developments by formulating an iterative process in which from an algebra equipped with defined addition, multiplication, vector multiplication, and conjugation operations, a new algebra with twice the dimension as the previous one is obtained as the Cartesian product of this algebra with itself. This process is known as the Cayley-Dickson construction. Thus, the construction over the field of real numbers yields a sequence of algebras of dimension $2^n$, leading to complex numbers ($n=1$), quaternions ($n=2$), octonions ($n=3$), sedenions ($n=4$), and so on. In physical sciences, hypercomplex numbers have been of great interest for modeling intrinsic properties of space-time in many applications such as dynamic vector modeling and Dirac equations. 

\subsection{Desiderata for hypercomplex numbers}
Starting from the two-dimensional real algebras complex numbers, consider $x = a + b\mathrm{i}\in \mathbb{C}  $ where $\mathrm{i}^2 = -1 $ is the imaginary part and $a,b \in \mathbb{R}$. In the two-dimensional algebras, there are also \textit{dual numbers} in which a nilpotent element is used in the basis as $x = a + b\epsilon$ where $\epsilon^2= 0$ and \textit{split-complex } where a hyperbolic imaginary unit is used leading $x = a + b\mathrm{j}$ where $\mathrm{j}^2 = 1$. This algebra satisfies the commutative, associative, and alternative properties. Following the Cayley-Dickson construction, the quaternion numbers are defined as $x = a + b\mathrm{i}+ c\mathrm{j}+ d\mathrm{k} \in \mathbb{H}$  where $a,b,c,d \in \mathbb{R}$ and $\mathrm{i}^2=\mathrm{j}^2=\mathrm{k}^2 = -1$. This algebra is no longer commutative but achieves associative and alternative properties. In this family of four-dimensional algebra, we also have split-quaternions in which $\mathrm{i}^2=-1,\mathrm{j}^2= {\mathrm{k}}^2=1$   which contains zero divisors, nilpotent elements, and nontrivial idempotents. Also, if coefficients $a,b,c,d \in \mathbb{C}$ i.e., $a=a_r + a_i\mathrm{I}, b=b_r + b_i\mathrm{I}, c=c_r + c_i\mathrm{I}, d=d_r + d_i\mathrm{I}$ where $\mathrm{I}^2=-1$  allows commutative property.

\begin{strip}
		\begin{tcolorbox}[colback=infocolor,title={}]
			\begin{table}[H]
				\centering
				\caption{Hypercomplex Representation and its Imaging Applications}
				\begin{tabular}{p{6cm} p{4.0cm} p{5.5cm}}
					\hline
					\textbf{Hypercomplex Number} & \textbf{Illustration$^{a,b}$}&\textbf{Image And Signal Processing Applications}\\
					\hline 
					\begin{minipage}{0.3\columnwidth}
                    \textcolor{blue}{\textbf{Complex Number}}: $x = a + b \mathrm{i} \in \mathbb{C}$ 
                    
                    where $\mathrm{i}^2 = -1, a,b \in \mathbb{R}$. 
                    
                    \begin{table}[H]\centering
\resizebox{1\textwidth}{!}{\begin{tabular}{ccc}
\multicolumn{3}{c}{\textbf{Algebra properties}}\\\toprule[1pt]
\multicolumn{3}{c}{$x, y, z \in \mathbb{C}$}\\
\textbf{Commutative} & \textbf{Associative}                &               \textbf{Alternative} \\
$xy=yx$ & $(xy)z = x(yz)$ & $(yx)x=y(xx)$\\
\checkmark & \checkmark &\checkmark
\\ \toprule[1pt]
\end{tabular}}
\end{table}
					\end{minipage} & \begin{minipage}{0.4\columnwidth}
                    \includegraphics[width=0.5\linewidth]{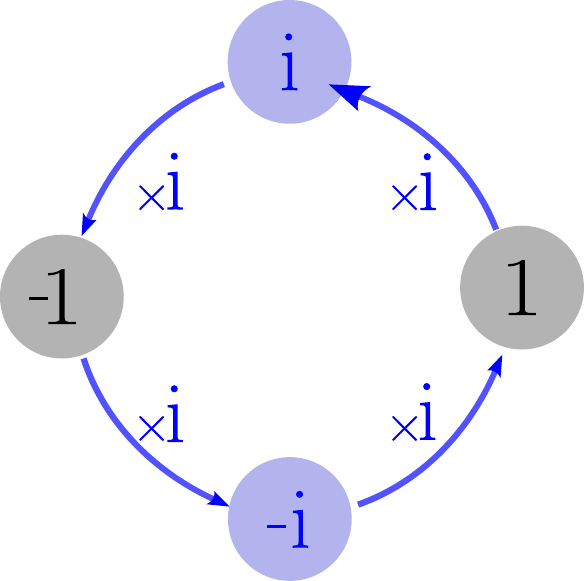}
					\end{minipage}& \begin{minipage}{0.3\columnwidth}
     Traditional image and signal processing methods employ complex-valued signals, where the flagship examples are in communications constellation coding and optical imaging PR.. 
					\end{minipage}\\
					\hline
		\begin{minipage}{0.3\columnwidth}
                    \textcolor{blue}{\textbf{Quaternion}}: $x = a + b \mathrm{i} + c \mathrm{j} + d \mathrm{k} \in \mathbb{H}$ 
                    
                    where $\mathrm{i}^2=\mathrm{j}^2=\mathrm{k}^2= {-1}, a,b,c,d \in \mathbb{R}$. 
                    
                    \begin{table}[H]\centering
\resizebox{1\textwidth}{!}{\begin{tabular}{ccc}
\multicolumn{3}{c}{\textbf{Algebra properties}}\\\toprule[1pt]
\textbf{Commutative} & \textbf{Associative}                &               \textbf{Alternative} 
\\\textcolor{red}{X} & \checkmark &\checkmark
\\ \toprule[1pt]
\end{tabular}}
\end{table}
					\end{minipage} & \begin{minipage}{0.4\columnwidth}
                    \includegraphics[width=0.5\linewidth]{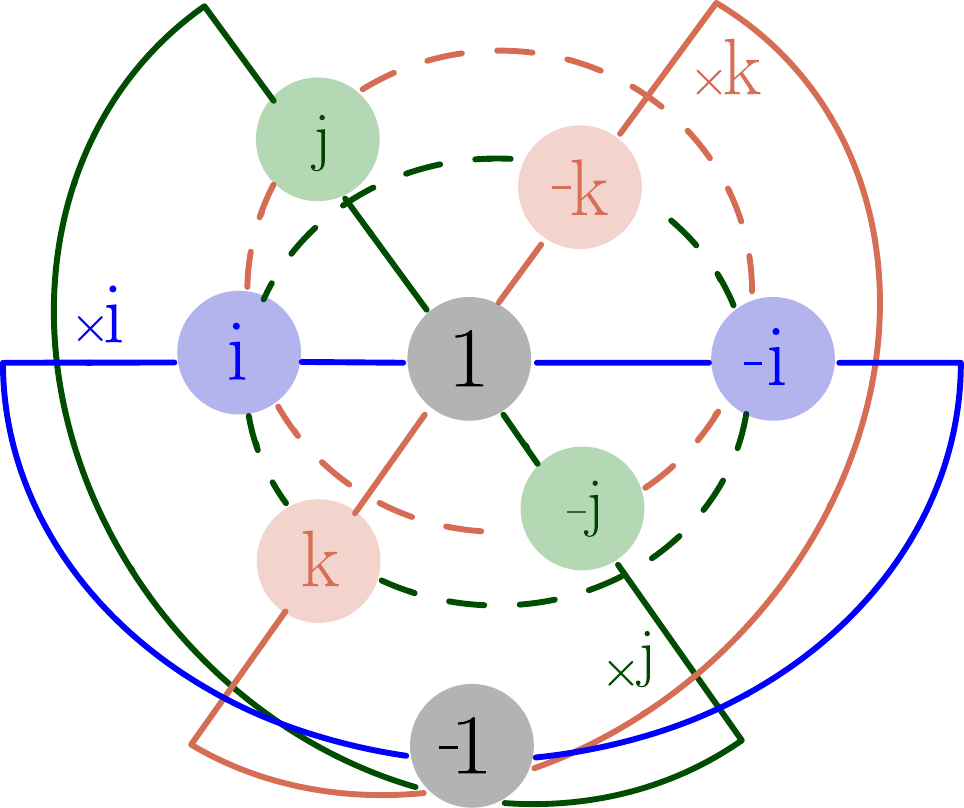}
					\end{minipage}& \begin{minipage}{0.3\columnwidth}
     Quaternion representation has been widely used in RGB image processing applications by harnessing the correlation between color channels via quaternion algebra. Some examples are PCA, matrix completion, low-rank approximation, and PR.
					\end{minipage}\\
					\hline
		\begin{minipage}{0.3\columnwidth}
                    \textcolor{blue}{\textbf{Split-Quaternion}}: $x = a + b \mathrm{i} + c \mathrm{j} + d \mathrm{k} \in \mathbb{H}_s$ 
                    
                    where $\mathrm{i}^2=-1, \mathrm{j}^2=\mathrm{k}^2= 1, a,b,c,d \in \mathbb{R}$. 
                    
                    \begin{table}[H]\centering
\resizebox{1\textwidth}{!}{\begin{tabular}{ccc}
\multicolumn{3}{c}{\textbf{Algebra properties}}\\\toprule[1pt]
\textbf{Commutative} & \textbf{Associative}                &               \textbf{Alternative} 
\\\textcolor{red}{X} & \checkmark &\checkmark
\\ \toprule[1pt]
\end{tabular}}
\end{table}
					\end{minipage} & \begin{minipage}{0.4\columnwidth}
                    \includegraphics[width=0.5\linewidth]{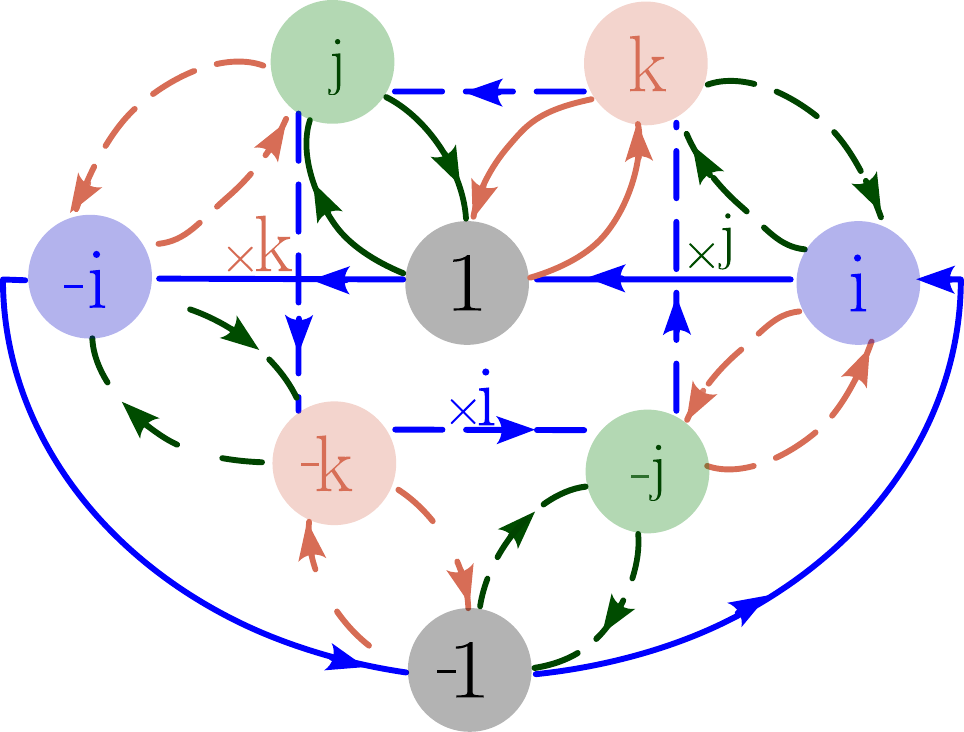}
					\end{minipage}& \begin{minipage}{0.3\columnwidth}
     In the context of signal processing, split-quaternion formulations have been proposed in adaptive filtering design, neural networks, and RGB image processing. The split-quaternion approaches handle noncommutative properties that quaternion methods ignore. 
					\end{minipage}
     \\
					\hline
		\begin{minipage}{0.3\columnwidth}
                    \textcolor{blue}{\textbf{Biquaternion}}: $x = a + b \mathrm{i} + c \mathrm{j} + d \mathrm{k}\in \mathbb{H}_b$ 
                    
                     $\mathrm{i}^2= \mathrm{j}^2=\mathrm{k}^2=\mathrm{I}^2=- 1,a,b,c,d \in \mathbb{C}$. 

                    $a = a_r + a_i \mathrm{I},b = b_r + b_i \mathrm{I},c = c_r + c_i \mathrm{I},d = d_r + d_i \mathrm{I}$
                    \begin{table}[H]\centering
\resizebox{1\textwidth}{!}{\begin{tabular}{ccc}
\multicolumn{3}{c}{\textbf{Algebra properties}}\\\toprule[1pt]
\textbf{Commutative} & \textbf{Associative}                &               \textbf{Alternative} 
\\\checkmark & \checkmark &\checkmark
\\ \toprule[1pt]
\end{tabular}}
\end{table}
					\end{minipage} & \begin{minipage}{0.4\columnwidth}
                    \includegraphics[width=0.5\linewidth]{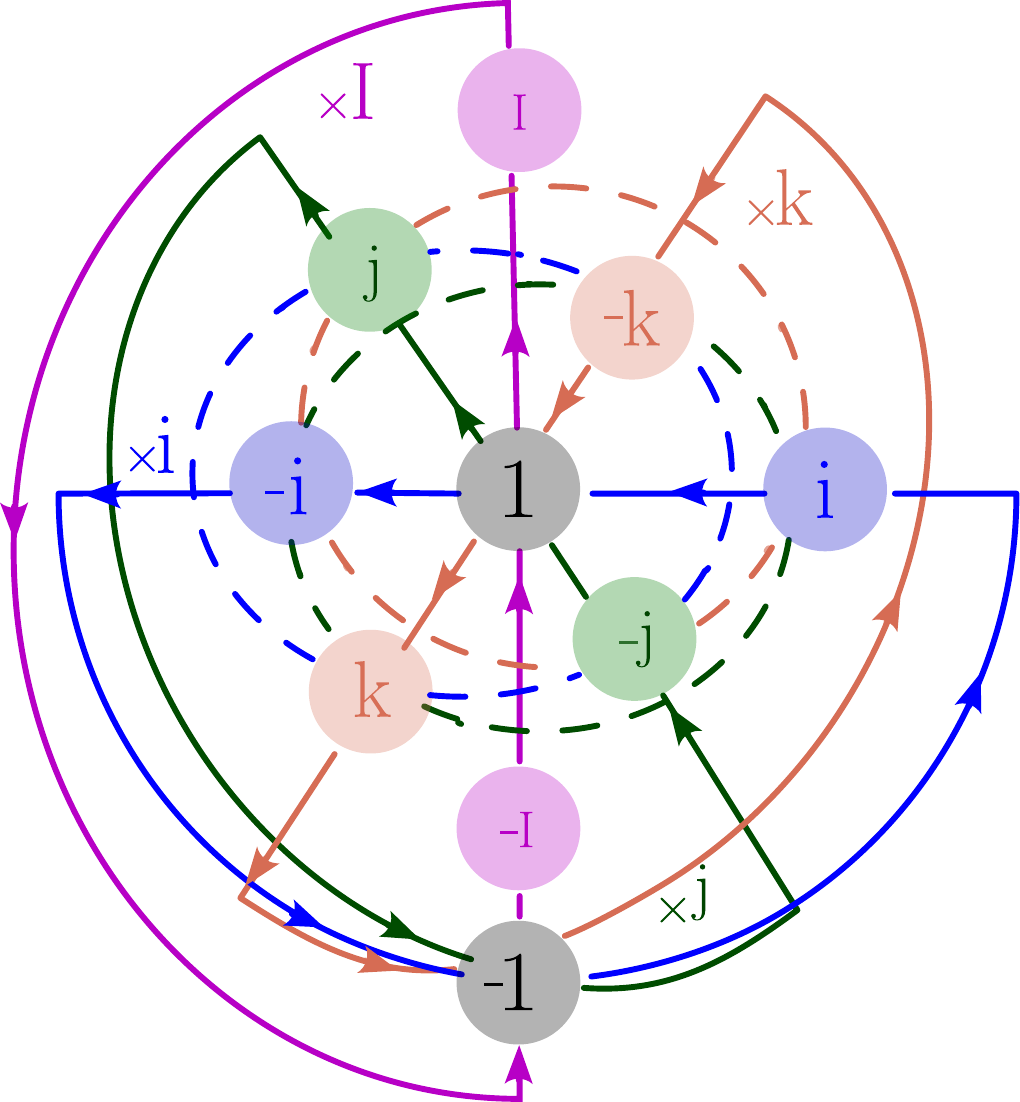}
					\end{minipage}& \begin{minipage}{0.3\columnwidth}
     Biquaternions are generalized quaternion numbers with complex-valued coefficients. This algebra admits multiplication commutative property which has been found useful in color image processing, array sensor processing, and convolutional neural networks.
					\end{minipage}
     \\
					\hline
		\begin{minipage}{0.3\columnwidth}
                    \textcolor{blue}{\textbf{Octonion}}:{ $x = a_0+\sum_{i=1}^7 a_i \mathrm{e}_i\in \mathbb{O}$ 
                    
                     $\mathrm{e}_i^2=-1, 1\leq i \leq 7 ,a_i \in \mathbb{R}$
                    }\begin{table}[H]\centering
\resizebox{1\textwidth}{!}{\begin{tabular}{ccc}
\multicolumn{3}{c}{\textbf{Algebra properties}}\\\toprule[1pt]
\textbf{Commutative} & \textbf{Associative}                &               \textbf{Alternative} 
\\\textcolor{red}{X}& \textcolor{red}{X} &\checkmark
\\ \toprule[1pt]
\end{tabular}}
\end{table}
					\end{minipage} & \begin{minipage}{0.4\columnwidth}
                    \includegraphics[width=0.5\linewidth]{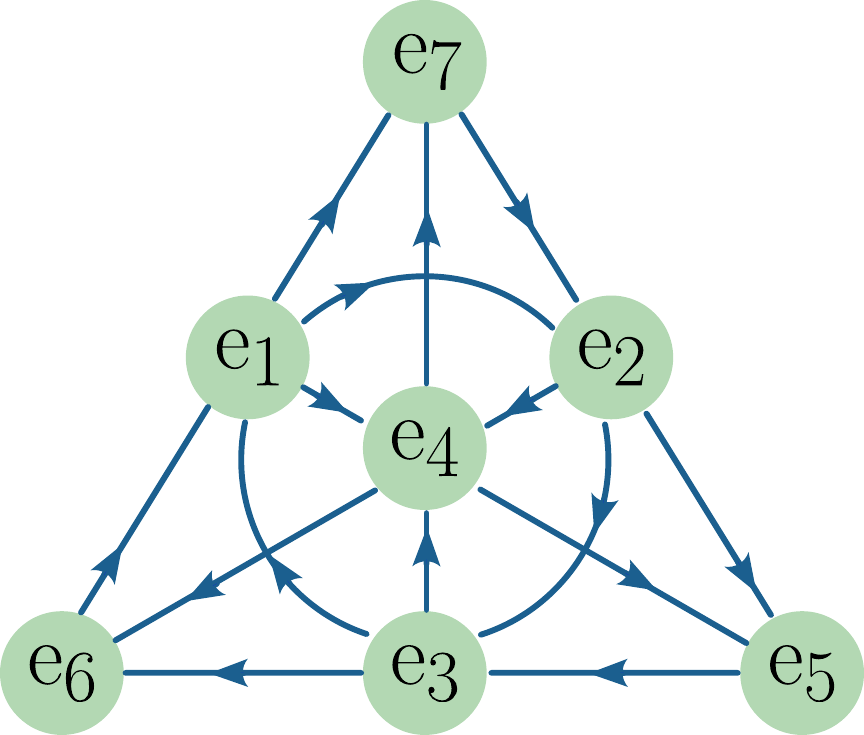}
					\end{minipage}& \begin{minipage}{0.3\columnwidth}
     Octonions generalize quaternion algebra to 8-dimensional signals via Cayley–Dickson construction. Octonion representation has been widely employed for color-stereo image processing, multispectral image restoration, deep neural networks, and PR.
					\end{minipage}\\\toprule[1pt]
				\end{tabular}
				\label{tab:1}
			\end{table}
   $^{a}$The graphs depict the multiplication properties of each algebra and the relationship among different imaginary units. The arrow color indicates the imaginary unit multiplied with the node at the beginning of the arrow to yield the node at the end of the arrow. 
   {$^b$ Dotted lines denote the product lines that lead to every imaginary unit while solid lines lead to $\pm1$. }
		\end{tcolorbox}
\end{strip}
Finally, an octonion number $x$ is defined as {$x=a_0+\sum_{i=1}^{7} a_i\mathrm{e}_i$}, where $a_i$ are real-valued coefficients and $\mathrm{e}_i$ are the \textit{octonion units} such that $\mathrm{e}_i^2=-1$ for $i=1,\dots,7$. Octonion algebra is non-associative and non-commutative. Table \ref{tab:1} summarizes these properties and shows the product properties of each algebra, along with the most relevant applications in signal and image processing \subsection{Hypercomplex signal processing}
In signal processing, finding appropriate modeling, analysis, and processing tools for high-dimensional signals is an ongoing challenge. Over the last three decades, HSP has emerged as a highly suitable approach to high-dimensional signal processing by exploiting the corresponding algebra to holistically process all signal dimensions jointly instead of independent dimension processing as in traditional real/complex approaches. A major area of application of HSP is image processing, where color images are represented as a \textit{quaternion} (with the red-green-blue pixel at any location encoded as a quaternion-valued function) thereby allowing holistic processing of the image instead of dealing with each color channel independently. Quaternion image representation is essential to improved performance in several applications, including color-sensitive image filtering through hypercomplex convolution, correlation techniques, color edge detection, and principal component analysis (PCA)~{\cite{miron2023quaternions,chen2019low}}.  

 Further, in applications such as image painting and edge detection \cite{pei2004commutative}, \textit{biquaternion} (a modified quaternion that enables commutative multiplication) is employed. Using Cayley–Dickson algebra, quaternions are extended to eight-dimensional \textit{octonions},
 with promising applications in multispectral imagining (MSI) processing (seven-color channel images), wherein each channel is assigned to a different imaginary component of an octonion number. In particular, octonions have been used for MSI denoising via dictionary learning. Octonion representations are also useful for image watermarking and image reconstruction with color stereo images, where two RGB images are staked as a six-layer image, where each layer is defined as a different imaginary component of the octonion {\cite{lazendic2018octonion,yamni2021novel}}.

 Similar to real- or complex-valued signals, HSP allows spectral representation of quaternions and octonions via their Fourier transforms (FT) \cite{ell2014quaternion}. A quaternion/octonion FT (QFT/OFT) allows the transformation of quaternion-valued signals from their space-time domain to the quaternion frequency domain. HSP using these transforms is widely used in color image filtering and motion estimation. Other HSP tools have also been developed recently, including the short-time Fourier transform (STFT), fractional Fourier transform (FrFT), wavelet transform (WT), linear canonical transform (LCT), bispectrum analysis, and deep learning. These developments are important for PR sensing modalities because the ill-posedness of the conventional PR problem is often addressed through additional measurements (e.g., Fourier, STFT, FrFT, or wavelet) \cite{pinilla2021wavemax,pinilla2022phase}. 
 
\section{Conventional Complex-Valued Phase Retrieval}
Conventional PR has been significantly researched signal processing problem, where we want to recover a complex-valued signal $\bsym x \in \mathbb{C}^{n}$  given its amplitude-only measurements $\mathbf y\in \mathbb{C}^{m}$ as $\bsym y = \vert\mathbf{A x }\vert^2 $, where the known measurement matrix $\mathbf A \in \mathbb{C}^{m\times n}$.  This problem arises in several areas such as diffractive imaging \cite{pinilla2023unfolding}, X-ray crystallography \cite{pinilla2018phase}, astronomy \cite{fienup1987phase},
and radar waveform design \cite{pinilla2021banraw}. According to the physical application of the PR formulation, the sensing matrix $\mathbf{A}$ is defined accordingly. For instance, in near-field optical imaging \cite{pinilla2023unfolding}, the sensing matrix is the Fourier matrix, and a diffractive optical element (DOE) is employed to reduce the ambiguity of the phaseless Fourier measurements. Usually, to improve the recovery performance, multiple snapshots are acquired varying the DOE. In this case, the acquisition model becomes $\mathbf{y} = \vert\left[\mathbf{D}_1^H\mathbf{F}^H,\dots,\mathbf{D}_L^H\mathbf{F} ^H\right]^H\mathbf{x}\vert^2$ where $\mathbf{F}\in \mathbb{C}^{n\times n}$  is the discrete Fourier transform (DFT) and $\mathbf{D}_\ell \in \mathbb{C}^{n\times n}$ for $\ell = 1,\dots, L$ are diagonal matrices whose entries are the diagonalization of the DOE coding elements and $L$ is the number of acquired snapshots.  

In other applications, for instance, in measuring ultrashort laser pulses \cite{trebino1993using}, the phaseless measurements are obtained via short-time Fourier transform (STFT), where a window $\mathbf{w}$ with compact frequency support of size $R$, i.e., $\Vert \mathbf{w}\Vert_\infty = R$.  Then, the measurements are $\mathbf{y} = \vert \left[\mathbf{W}_1 ^H\mathbf{F}^H,\dots,\mathbf{W}_L^H \mathbf{F}^H\right]^H\mathbf{x}\vert^2$ where $\mathbf{W}_\ell$  is a diagonal matrix with entries given by the $L-$circular shifting of the window $\mathbf{w}$. Thus, multiple sliding window measurements reduce the ill-posedness of the PR problem. Beyond imaging applications, other formulations have been employed, for instance, in audio processing, the problem of recovering phase information after computing the modulus of the wavelet transform (scalogram) is of wide interest \cite{waldspurger2017phase}. Here, the sensing becomes $\mathbf{y} = \vert\left[\bsym{\Phi}_1^H, \dots \bsym\Phi_L^H\right]^H\mathbf{x}\vert^2 $, where $\bsym\Phi_\ell$ is a convolution matrix built upon {dilated} version of the mother wavelet $\bsym{\phi}_\ell$. 

The PR recovery generally aims to solve the following optimization problem:
\begin{equation}
    \minimize_{\widetilde{\mathbf{x}}\in\mathbb{C}^n} \frac{1}{2m} \sum_{\ell=1}^{m}\mathcal{L}(y_\ell,\vert \langle \widetilde{\mathbf{x}},\mathbf{a}_\ell\rangle \vert^2)\label{eq:conv_pr}
\end{equation}
where $\mathbf{a}_\ell$ is the $\ell$-th row of the sensing matrix and $\mathcal{L}$ is a fidelity measure. Due to the PR model, solving \eqref{eq:conv_pr} includes a factor ambiguity of phase factor $e^{\mathrm{i}\theta}, \theta \in [0,2\pi]$, since $\vert \mathbf{x}\vert =\vert \mathbf{x}e^{\mathrm{i}\theta}\vert$. Thus, the recovery distance is defined as $d(\widetilde{\mathbf{x}},\mathbf{x}) = \min_{\theta}\Vert \mathbf{x}e^{\mathrm{i}\theta}-\widetilde{\mathbf{x}}\Vert_2$. An $\epsilon$-feasible solution of \eqref{eq:conv_pr} is defined by the region $E_\epsilon(\widetilde{\mathbf{x}}) = \{\widetilde{\mathbf{x}}\in \mathbb{C}^n:d(\widetilde{\mathbf{x}},\mathbf{x})<\epsilon\}$. 
\begin{table}[!t]
    \centering
    \caption{Quaternion and octonion identities}
    \begin{tabular}{ccc}\toprule[1pt]
        Definition & Quaternion & Octonion \\  & $x\in \mathbb{H},\mathbf{x}\in \mathbb{H}^n$ & $x\in \mathbb{O},\mathbf{x}\in \mathbb{O}^n$ \\\toprule[1pt]
        Conjugation & $x^* = a - b\mathrm{i}-c\mathrm{j}-d\mathrm{k}$ & $x^* =x_0  - \sum_{1}^{7}x_i\mathrm{e}_i$\\
        Modulus & $\vert {x} \vert = \sqrt{a^2+b^2+c^2+d^2}$  & $\vert {x} \vert = \sqrt{\sum_{i=0}^7 x_i^2}$\\
        Inverse & $x^{-1} = \frac{x^*}{\vert x \vert^{2}}$ & $x^{-1} = \frac{x^*}{\vert x \vert^{2}}$\\
       Vector Norm  & $\Vert \mathbf{x} \Vert = \sum_{i=0}^{n-1}\vert \mathbf{x}_i\vert$& $\Vert \mathbf{x} \Vert = \sum_{i=0}^{n-1}\vert \mathbf{x}_i\vert$\\
       \toprule[1pt]
    \end{tabular}
    \label{tab:definitions}
\end{table}
Considerable efforts have been devoted to the algorithm development (see, e.g., {\cite{vaswani2020nonconvex,pinilla2023unfolding,dong2023phase}} for recent surveys, and references therein). For example, convex optimization-based with extensive recovery theoretical guarantees {\cite{candes2013phaselift}} by converting the quadratic problem to a recovery of a rank-1 matrix via semidefinite programming. While high performance is achieved with these methods, the computational cost becomes infeasible at large-scale signal dimensions. On the other hand, non-convex optimization alleviates the computational complexity by performing gradient descent-like iterations. In these approaches, the selection of the fidelity function $\mathcal{L}$ plays a crucial role. For instance, the seminal paper on the {Wirtinger} flow (WF) algorithm \cite{candes2015phase_2} {employs a quadratic loss} of the squared magnitude measurements. {In \cite{chen2015solving}, the WF convergence is improved by changing the quadratic cost function for a Poisson-based data fidelity function}. Further improvements include using the least square function \cite{zhang2017nonconvex}, or median-truncated functions to obtain a robust algorithm to outliers \cite{zhang2018median}. 
 Another key component of these methods is the initialization algorithm, which is usually performed via spectral methods to compute the leading eigenvector of the measurement matrix.  Based on the vast PR literature, in the following section, we formulate HPR models and algorithms.

\section{Hypercomplex Phase retrieval}
Table \ref{tab:definitions} lists useful definitions for quaternion and octonion signals for which we now introduce PR formulations.

\subsection{Quaternion PR (QPR): Real-valued sensing matrix}
Consider estimating the quaternion signal $\mathbf{x}\in \mathbb{H}^n$ from quadratic intensity-only measurements $\mathbf{y} = \vert \mathbf{Ax}\vert^2$ where $\mathbf{A}\in\mathbb{R}^{m\times n}$ is the sensing matrix. This problem was first tackled in \cite{chen2022phase_2} and provides several characterizations to determine complex/vector-valued functions in a linear space of finite dimensions, up to a trivial ambiguity, from the magnitudes of their linear measurements. The vector-valued formulation was extended to {quaternion-valued functions}. Its analysis was developed also for affine PR of vector-valued functions, which is a formulation that is encountered in holography \cite{liebling2003local}. However, \cite{chen2022phase_2} did not propose a recovery algorithm.

 \begin{strip}
     \begin{tcolorbox}[colback=infocolor,title={\textbf{Quaternion PR}},boxsep=1pt,left=4pt,right=4pt,top=2pt,bottom=2pt]
	\begin{multicols}{2}
 \textbf{$\mathbb{HR}$ Calculus:} The QPR is based on the quaternion vector/matrix derivatives defined below. \vspace{3mm}

  \textit{Definition 1 (Quaternion Rotation) \cite{xu2015theory}}: is defined for any two quartenions $x$ and $\mu$, the transformation $$x^\mu \coloneqq  \mu x\mu ^{-1}$$
  is a 3-dimensional rotation of the vector part of $x$ about the vector part of $\mu$.
\vspace{3mm}

\textit{Definition 2  {(Generalized $\mathbb{HR}$ (GHR) derivatives)}\cite{xu2015theory}}: is defined for a  quaternion $x = a+b\mathrm{i}+c\mathrm{j}+d{\mathrm{k}}$, and a function $f$ with respect to the transformed quaternion $x^\mu$ as 
$$\frac{\partial_l f}{\partial x^{\mu}} = \frac{1}{4}\left(\frac{\partial f}{\partial a}-\frac{\partial f}{\partial b}\mathrm{i}^{\mu}-\frac{\partial f}{\partial c}\mathrm{j}^{\mu}-\frac{\partial f}{\partial c}\mathrm{k}^{\mu}\right).$$
(Without loss in generality, the $\mathbb{HR}$ calculus can be derived from either left or right GHR derivatives.)
\vspace{3mm}

\textit{Definition 3 (Quaternion gradient) \cite{xu2015theory}}: is defined for a scalar function $f(\mathbf{x})$, where $\mathbf{x}\in\mathbb{H}^{n}$, the gradient of $f$ with respect to $\mathbf{x}$ is defined as
$$\nabla_\mathbf{x} f = \left(\frac{\partial f}{\partial \mathbf{x}^\mu}\right)^T,$$ 
where $\frac{\partial f}{\partial \mathbf{x}^\mu} = \left[\frac{\partial f}{\partial \mathbf{x}_1^{\mu}}, \dots, \frac{\partial f}{\partial \mathbf{x}_n^{\mu}}\right]$.

\textbf{Gradient of QPR cost function}
In conventional PR, WF performs gradient-descent-like iterations based on the complex-valued calculus denoted by $\mathbb{CR}$  (also known as the Wirtinger calculus). In QPR, to compute QWF iterations of the cost function in \eqref{eq:qpr_opt}, GHR derivatives are applied as
\begin{align}
    \frac{\partial f(\widetilde{\mathbf{x}})}{\partial\widetilde{\mathbf{x}}}& = \frac{1}{2m} \sum_{\ell=1}^{m} \frac{\partial}{\partial\widetilde{\mathbf{x}}} \left(\vert\mathbf{a}_\ell^*\widetilde{\mathbf{x}}\vert^4- 2y_\ell\vert\mathbf{a}_\ell^*\widetilde{\mathbf{x}}\vert^4 \right)\nonumber\\&=\frac{1}{m} \sum_{\ell = 1}^{m}(\vert \mathbf{a}^*_\ell \widetilde{\mathbf{x}}\vert^2-y_\ell)  \widetilde{\mathbf{x}}^*\mathbf{a}_\ell\mathbf{a}_\ell^*
\end{align}
Figure~\ref{fig:transition_qpr}a plots the performance of QWF by employing $\mathbb{HR}$ calculus. where the sample complexity $m/n$ was varied from 100 Monte-Carlo simulations. The reconstructions with real RGB image data are shown in Figure~\ref{fig:transition_qpr}b. Results suggest that QWF allows a better performance for both synthetic signals and real RGB signals.
			\begin{center}
			\includegraphics[width=1.0\linewidth]{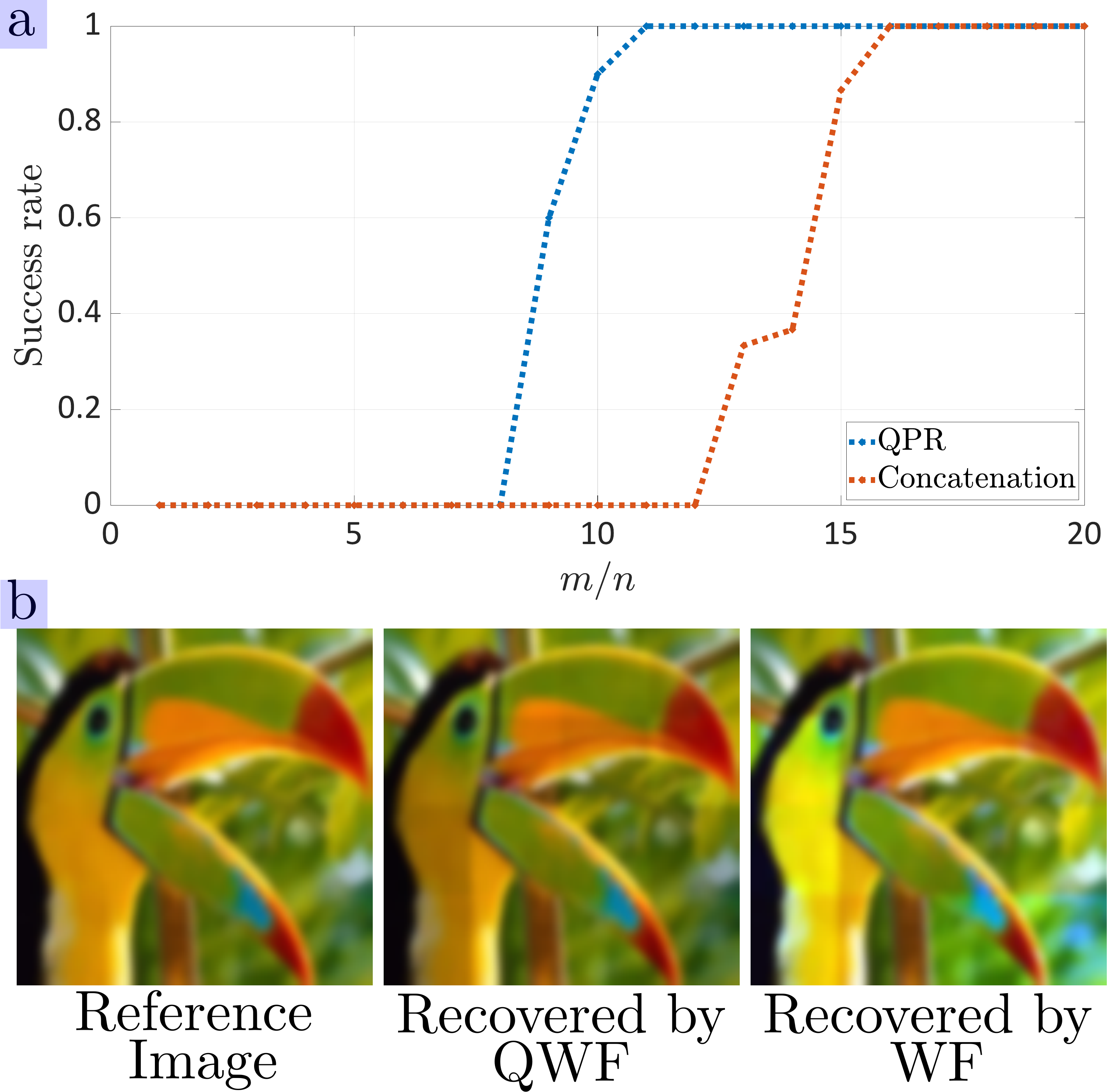}
				\captionof{figure}{(a) Recovery performance of QWF for different sample complexity and WF algorithm by concatenating the four dimensions of the quaternion signal. {For the quaternion setting, the sensing matrix was drawn from a quaternion normal distribution. For the concatenation, the same was drawn from a complex normal distribution}. The experiments are performed for $n=500$. (b) Illustration of QPR recovery using a red-green-blue (RGB) image. Here, we set each color channel as the imaginary parts of a quaternion signal and reconstruct it with QWF. For comparison, by concatenating each color channel and applying WF, the recovery is poorer than QPR. In both cases, reconstruction was performed in patches of $N=32$ such that $n=1024$ and $m/n = 15$.} 
                \label{fig:transition_qpr}       
			\end{center}
        \end{multicols}
        \vspace{2mm}
    \end{tcolorbox}
\end{strip}    

However, the sensing matrix in this problem is limited to only real values, where the model fails to harness the quaternion multiplication. Instead, it considers $\mathbb{H}$ as $\mathbb{R}^4$ space, thereby reducing QPR to a special case of PR of a real vector-valued signal. Moreover, $\mathbf{A} \in \mathbb{R}^{m\times n}$ leads to more trivial ambiguities that may limit applications of QPR.

\subsection{Quaternion PR: Quaternion-valued sensing matrix} \label{sec:qwf}
In general, HPR entails the estimation of a hypercomplex number measured through a hypercomplex sensing matrix. In this context, unlike \cite{chen2022phase_2}, \cite{chen2023phase} considered a quaternion-valued sensing matrix $\mathbf{A}\in \mathbb{H}^{m\times n}$ to solve the following optimization problem 
\begin{equation}    \minimize_{\widetilde{\mathbf{x}}\in\mathbb{H}^n}f(\mathbf{\widetilde{x}}) =  \frac{1}{2m} \sum_{\ell=1}^{m}(\vert \mathbf{a}_\ell^*\mathbf{\widetilde{x}}\vert^2-y_\ell)^2.\label{eq:qpr_opt}
\end{equation}

To tackle \eqref{eq:qpr_opt}, \cite{chen2023phase} proposed quaternion  WF (QWF) that comprises an initialization stage followed by the recovery. The initialization estimates $\widetilde{\mathbf{x}}_0$, which has an approximated direction of $\mathbf{x}$.  Then, a spectral method yields the leading eigenvector $\mathbf{v}$ of the Hermitian matrix $\mathbf{Y} = \frac{1}{m}\sum_{\ell=1}^{m}y_\ell \mathbf{a}_\ell \mathbf{a}_\ell^* \in\mathbb{H}^{n\times n}$; note that $\mathbb{E}[\mathbf{Y}] = \mathbf{I}_n + \frac{1}{2}\mathbf{xx}^*$ {where the expectation is {with} respect to the random rows of the sensing matrix}. Since $\mathbb{E}[y_\ell] = \Vert\mathbf{x}\Vert_2^2$, the \textit{scaling factor} $r = (\frac{1}{m}\sum_{\ell=1}^{m} y_\ell^2)^{1/2}$ is approximately the magnitude of $\mathbf{x}$ when $m$ is sufficiently large. The initial value becomes $\mathbf{\widetilde{x}}^0 = r\mathbf{v}$. To derive the QWF iteration, we rely on the quaternion calculus $\mathbb{HR}$ to perform the derivatives of the cost function \eqref{eq:qpr_opt}.

Then, QWF employs GHR  to compute gradient-descent-like iterations (see Quaternion PR box):
\begin{equation}
    \widetilde{\mathbf{x}}^{i+1} =\widetilde{\mathbf{x}}^{i} - \eta \nabla_{\widetilde{\mathbf{x}}}f({\widetilde{\mathbf{x}}}^i),
\end{equation}
for $i=1,\dots,I$ iterations. This algorithm has been proven to converge at linear rates under the distance, \begin{equation}
d(\mathbf{x},\widetilde{\mathbf{x}}) = \min_{{w}} \Vert\widetilde{\mathbf{x}}-\mathbf{x}w\Vert_2 = \Vert\widetilde{\mathbf{x}}-\mathbf{x}\operatorname{sign}(\mathbf{x}^*\widetilde{\mathbf{x}})\Vert_2,
\end{equation} where $\vert{w}\vert = 1$ is a constant phase factor and $\operatorname{sign}(x) = \frac{x}{\vert x \vert}$, when the quaternion components of the signal are independent~\cite{chen2023phase}. 

Other WF-like variations of QPR also exist. For instance, quaternion truncated WF (QTWF) changes the cost function in \eqref{eq:qpr_opt} to $f(\widetilde{\mathbf{x}}) = \frac{1}{m}\sum_{\ell=1}^{m}y_\ell\log(\vert\mathbf{a}_\ell^*\widetilde{\mathbf{x}}\vert)-\vert\mathbf{a}_\ell^*\widetilde{\mathbf{x}}\vert $ and performs GHR to develop the gradient descent iterations. In this case, the gradient is trimmed to be more selective by truncation. Similarly, the QWF variants for base truncated amplitude WF \cite{pinilla2023unfolding} 
and pure quaternion signal has also been defined \cite{chen2023phase}. 

A different QPR problem in \cite{shao2014double} considers the quaternion gyrator domain for a double-image encryption application. This transformation is a linear canonical integral transform which is viewed as an extension 
\noindent of Fourier analysis. It admits two quaternion signals $\mathbf{x}_1$ and $\mathbf{x}_2$ such that $\mathbf{y} = \vert \mathbf{A}_1 \mathbf{x}_1\vert^2= \vert \mathbf{A}_2 \mathbf{x}_2\vert^2$  where $\mathbf{A}_1$ and $\mathbf{A}_2$ are built upon \textit{gyrator transform} \cite{shao2013quaternion} with different rotation angles and phase mask that perform the encryption of the images. 

\subsection{Octonion PR (OPR)}
Ocotnions have garnered attention in applications including MSI processing \cite{lazendic2018octonion}, wherein each pixel 7-color channel image has a vector-valued representation such that each channel corresponds to different complex-variable dimensions. Octonions are employed for color-stereo image analysis to represent two 3-color channel images in a different imaginary dimension. The mutual processing along the color channels and two stereo images has been shown to improve the analysis. Recently, there has been broad interest in the recovery of the complex MSI represented using octonion-valued signals from its phaseless measurements. 

Consider the octonion-valued signal $\bsym{x} \in \mathbb{O}^{n}$ and its phaseless measurements  $\mathbf{y} = \vert\mathbf{A}\mathbf{x}\vert^2\in \mathbb{R}_+^m$ where $\mathbf{A}\in\mathbb{O}^{m\times n}$ is the octonion-valued sensing matrix. Note that, given a unit octonion $q$ such that $\vert q\vert = 1$, the signal $\mathbf{x}$ scaled by a global right pure octonion factor leads to the same $\mathbf{y}$, i.e., $\vert\mathbf{Ax}q\vert^2=\vert\mathbf{Ax}\vert^2$. The octonion algebra is non-commutative and, hence, $\vert \mathbf{A}q\mathbf{x}\vert^2\neq\vert\mathbf{Ax}\vert^2$. Then, OPR aims to recover $\mathbf{x}$ up to a \textit{trivial ambiguity} of only on the right octonion phase factor \cite{jacome2023octonion}. 

As before, the octonion WF (OWF) \cite{jacome2023octonion} employs the spectral initialization, wherein our goal is to obtain the initial estimate of the true signal by computing the leading eigenvector of the octonion-valued matrix $\mathbf{Y} = \frac{1}{m}\sum_{\ell=1}^{m} {{y}_\ell} \mathbf{a}_\ell\mathbf{a}_\ell^* \in \mathbb{O}^{n\times n}$. This may be achieved by solving an octonionic right eigenvalue decomposition. This decomposition has been solved earlier for small octonion-valued matrices ($n<4$) but it cannot be extended to larger matrices. Therefore, \cite{jacome2023octonion} adapted the power method for the right quaternion eigenvalue decomposition to compute the leading eigenvalue of $\mathbf{Y}$. This method relies on the real-matrix representation of octonion numbers (see Octonion PR box). Using power iterations over the real matrix representation, the inverse real representation $\aleph^{-1}(\cdot)$ is employed to yield the equivalent octonion leading eigenvalue. 

The octonion algebra is also non-associative. Hence, it lacks a clear definition of octonion derivatives, including chain rule, high-dimensional gradients, and gradient-based methods such as the WF.  Optimization methods that employ octonion representation (e.g., in singular value decomposition (SVD) or deep octonion neural networks) usually resort to the real-matrix representation to perform optimization over the real-valued variable.

 Inspired by this approach, \cite{jacome2023octonion} employed this representation to solve the following OPR problem:
\begin{align}
\minimize_{\widetilde{\mathbf{x}}\in \mathbb{O}^{n}} \sum_{\ell=1}^{m} {\frac{1}{2}}\left(\vert\mathbf{a}_\ell^* \widetilde{\mathbf{x}}\vert^2- {y}_\ell\right)^2.
\end{align}
Employing the real matrix representation, OPR is formulated as
\begin{align}
    \widetilde{\mathbf{x}} &= \aleph^{-1}\left(\argmin_{{\mathbf{x}}\in \mathbb{R}^{8n}} \sum_{\ell=1}^{m}  {\frac{1}{2}}\left(\Vert\gimel\left(\mathbf{a}_\ell^*\right)\aleph({\mathbf{x}})\Vert_2^2- {y}_\ell\right)^2\right)\nonumber\\&=  \aleph^{-1}\left(\argmin_{{\mathbf{x}}\in \mathbb{R}^{8n} }f({\mathbf{x}}) \right),\label{eq:problem_real}
\end{align}
where the $\ell_2$ norm comes from the observation that the norm of an octonion variable is the norm of its real representation. Then, \eqref{eq:problem_real} is solved by applying the gradient descent steps on the cost function as 
\begin{align}
\nabla_{ {\mathbf{x}}} f( {\mathbf{x}}) = \sum_{\ell=1}^{m}(\Vert\gimel\left(\mathbf{a}_\ell^*\right)\aleph( {\mathbf{x}})\Vert_2^2-\mathbf{y}_\ell)\gimel\left(\mathbf{a}_\ell^*\right)^T\gimel\left(\mathbf{a}_\ell^*\right)\aleph( {\mathbf{x}}).
\end{align}
 Then, the OWF update process in the $i$-th iteration, where $i\in \{1,\dots,I\}$ such that $I$ is the maximum number of iterations, becomes
\begin{equation}
    \widetilde{\mathbf{x}}^{(i)} = \widetilde{\mathbf{x}}^{(i-1)} - \alpha \nabla f(\widetilde{\mathbf{x}}^{(i-1)})
\end{equation}
where $\alpha$ is a suitable selected gradient step size. Finally, we obtain the octonion signal by computing the inverse real representation of $\widetilde{\mathbf{x}}^{(I)}$ as $\widetilde{\mathbf{x}} = \aleph^{-1} (\widetilde{\mathbf{x}}^{(I)})$.

{
Each quaternion (octonion) multiplication requires 16 (64) real multiplications \cite{cariow2020algorithm}. The standard WF algorithm has $\mathcal{O} (mn^2\log (1/\epsilon))$ complexity to achieve an accuracy of $d(\mathbf{x},\widetilde{\mathbf{x}}) \leq \epsilon \Vert \mathbf{x} \Vert_2$ \cite{candes2015phase_2}. In comparison, the computational complexities of QWF and OWF are of the order $\mathcal{O} (16^3mn^2\log (1/\epsilon))$ and $\mathcal{O} (64^3mn^2\log (1/\epsilon))$, respectively. We refer the interested reader to \cite{cariow2023fast} for research on addressing the high computational complexity of quaternion and octonion signal processing techniques.
 }

 \section{Emerging HPR Applications}
 
Traditional HPR problems employ measurements using a sensing matrix drawn from a random distribution. However, analogous to the conventional PR, measurements may also be taken using other HSP tools such as hypercomplex Fourier or other matrices. We list some of these emerging applications below.
\subsection{Hypercomplex Fourier PR}
Recovering phase information of the light has a plethora of applications including object detection, microscopy, and

  \begin{strip}
    \begin{tcolorbox}[colback=infocolor,title={\textbf{Octonion PR }},boxsep=1pt,left=4pt,right=4pt,top=2pt,bottom=2pt]
	\begin{multicols}{2}
 \textbf{Octonion real matrix representation}
				Unlike quaternion algebra, a real-matrix representation of an octonion number does not exist. However, \cite{rodman2014hermitian} proposed a pseudo-real matrix representation that has been successfully employed by many octonion-valued signal  {applications. To obtain} To obtain this representation, define the real representation of the octonion number $x\in \mathbb{O}$ as $\aleph(x) = [x_0,x_1,\dots,x_7]^T\in \mathbb{R}^{8}$. Then, the injective mapping $\gimel : \mathbb{O} \rightarrow \mathbb{R}^{8\times 8}$ is the {real matrix representation} of an octonion number \cite{rodman2014hermitian}: 
\par\noindent\small
\begin{equation}
    \gimel(x) =\left[\begin{array}{rrrrrrrr}
x_0 & -x_1 & -x_2 & -x_3 & -x_4 & -x_5 & -x_6 & -x_7 \\
x_1 & x_0 & x_3 & -x_2 & x_5 & -x_4 & -x_7 & x_6 \\
x_2 & -x_3 & x_0 & x_1 & x_6 & x_7 & -x_4 & -x_5 \\
x_3 & x_2 & -x_1 & x_0 & x_7 & -x_6 & x_5 & -x_4 \\
x_4 & -x_5 & -x_6 & -x_7 & x_0 & x_1 & x_2 & x_3 \\
x_5 & x_4 & -x_7 & x_6 & -x_1 & x_0 & -x_3 & x_2 \\
x_6 & x_7 & x_4 & -x_5 & -x_2 & x_3 & x_0 & -x_1 \\
x_7 & -x_6 & x_5 & x_4 & -x_3 & -x_2 & x_1 & x_0
\end{array}\right].\nonumber
\end{equation}\normalsize
Both representations $\aleph$ and $\gimel$ are also easily extended to vector/matrix octonion variables {i.e.,  consider $\mathbf{A}\in\mathbb{O}^{m\times n}$, we have that $\aleph(\mathbf{A}) \in \mathbb{R}^{8n\times m}$ and $\gimel({\mathbf{A}})\in\mathbb{R}^{8m\times 8n}$. Consider $\mathbf{x} \in \mathbb{O}^{n}$ and $\mathbf{A} \in \mathbb{O}^{m\times n}$, it holds $\aleph(\mathbf{Ax}) = \gimel(\mathbf{A})\aleph(\mathbf{x})$ and $\Vert\mathbf{x}\Vert = \Vert \aleph(\mathbf{x})\Vert_2$.}

\textbf{OPR distance metric}
To measure the error between the estimated octonion signal and the ground truth signal, define the distance $d(\mathbf{x},\mathbf{x}^\ast) = \min_{z}\Vert \mathbf{x}^{\ast}-\mathbf{x}z\Vert$ where $z \in \{z \in \mathbb{O}\vert \vert z\vert = 1\}$ is only-phase octonion factor. We represent this distance in terms of the pseudo-real-matrix representation of octonions as $d(\mathbf{x}^\ast,\mathbf{x}) =  \min_{z}\Vert\aleph(\mathbf{x}^\ast)- \gimel(\mathbf{x}))\aleph(z)\Vert$, using the property $\Vert \mathbf{x}\Vert = \Vert \aleph(\mathbf{x})\Vert$. After some tedious algebra, we obtain $d(\mathbf{x},\mathbf{x}^{\ast}) = \Vert\aleph(\mathbf{x}^\ast)- \gimel(\mathbf{x})) g(\mathbf{x}^{\ast})\Vert$, where 
\begin{equation}
     g(\mathbf{x}^{\ast}) = \operatorname{sign}\left(\left(\gimel(\mathbf{x})^T\aleph(\mathbf{x}^{\ast})\right)\left(\gimel(\mathbf{x})^T\gimel(\mathbf{x}\right)^{-1}\right).
\end{equation}

Figure \ref{fig:visual} presents an application of OWF for MSI. Here, the spectral image (Figure \ref{fig:visual}a) from the CAVE multispectral image dataset was used by cropping it to $32\times 32$ pixels and selecting eight equispaced spectral bands from 31 original spectral bands ranging from 400 to 700 nm.  Figure~\ref{fig:snr} shows OWF recovery for noise measurements for different SNR values. 
			\begin{center}
			\includegraphics[width=0.95\linewidth]{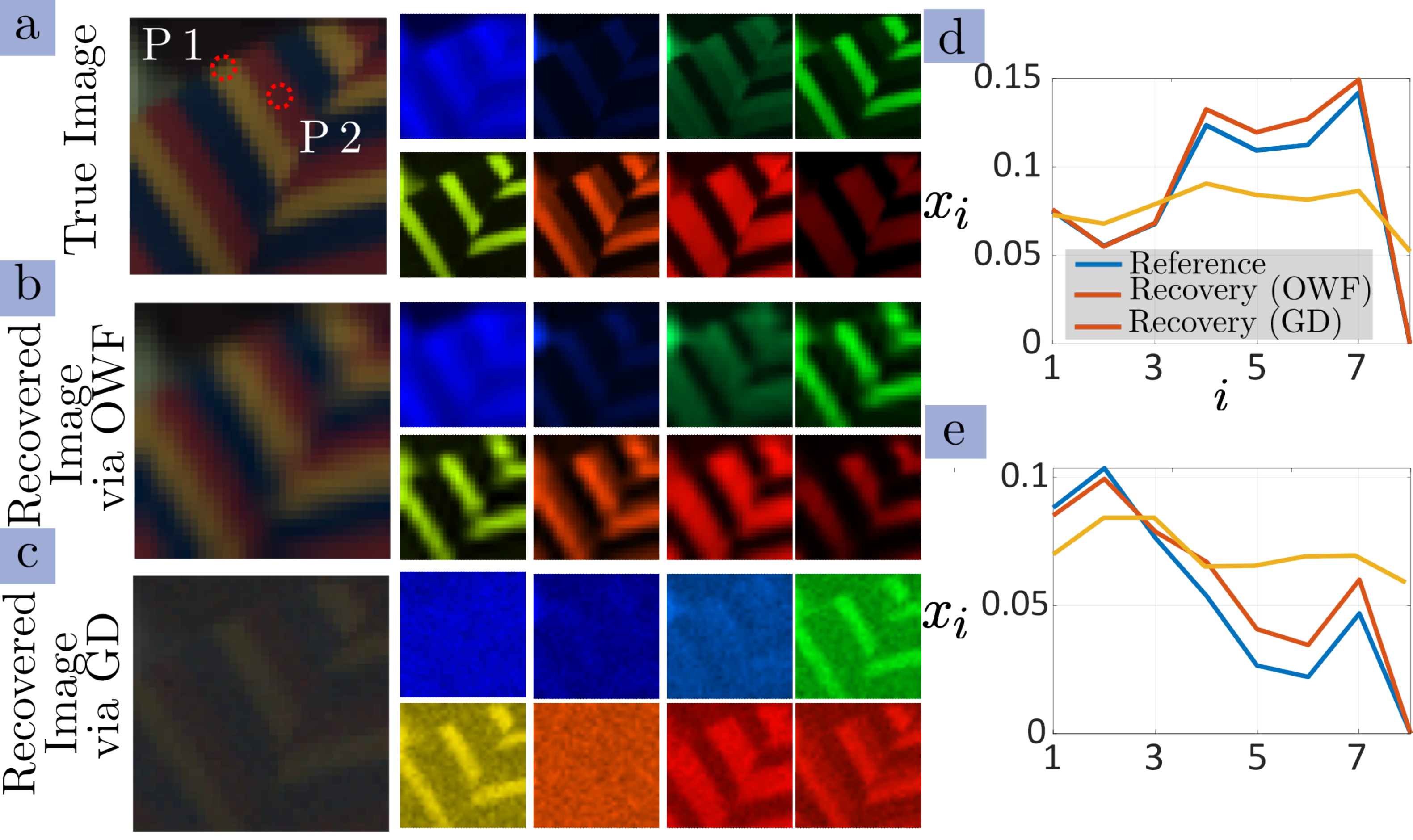}
				\captionof{figure}{Reconstruction of real data with OWF. (a) RGB representation of the eight-channel spectral image and its individual eight components on the right panel. (b) The OWF-reconstructed image and components. The PSNR of the recovered image is 39.007 [dB]. (c) The GD-based reconstructed image and components.  The PSNR of the recovered image is 24.1615 [dB]. Recovered octonion-valued numbers from two coordinates (c) (15,15) and (d) (10,10), labeled `P1' and `P2', respectively \cite{jacome2023octonion}. }
                \label{fig:visual}

                \includegraphics[width=0.95\linewidth]{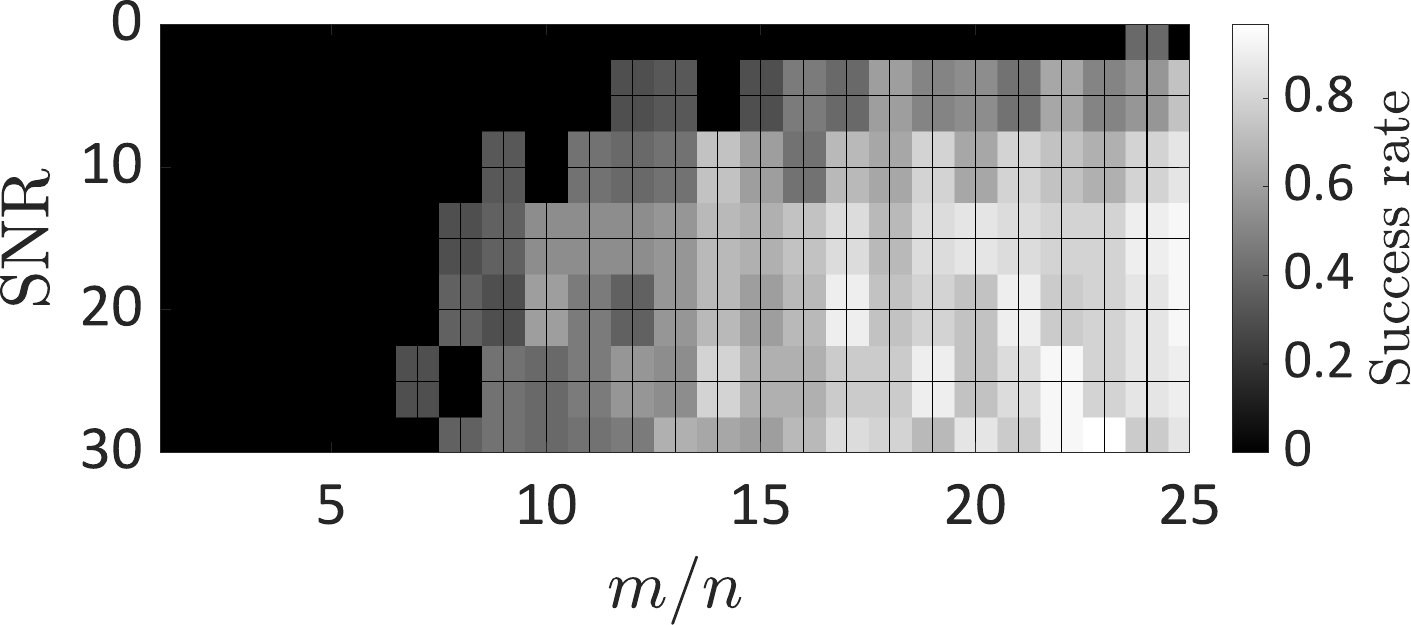}
				\captionof{figure}{Success rate of OWF for different values of sampling complexity $m/n$ with $n=30$ with measurements under additive Gaussian noise with a signal-to-noise ratio varying from 0 to 30 \cite{jacome2023octonion}.} \vspace{1cm}
				\label{fig:snr}
			\end{center}
        \end{multicols}
    \end{tcolorbox}
\end{strip}

\begin{strip}
		\vspace{-2em}
		\begin{tcolorbox}[colback=infocolor,title={\textbf{Fourier Quaternion PR in Optical Imaging}},boxsep=1pt,left=4pt,right=4pt,top=2pt,bottom=2pt]
			\begin{multicols}{2}
\textit{Quaternion Fourier Transform (QFT)} 
 \cite{ell2014quaternion}  {In quaternion signal processing, the QFT { does not have a unique definition} due to the position of the exponential or the pure quaternion unit employed}. For Fourier QPR, we employ the 2-D QFT (or two-sided QFT). Consider the continuous-valued signal $f(\mathbf{x}): \mathbb{R}^2\rightarrow \mathbb{H}, \mathbf{x} = [x_1,x_2] \in \mathbb{R}^2$, whose QFT $F(\mathbf{u})=\mathcal{F}_Q\{f(\mathbf{x})\}$, where $F(\cdot)$ is a function of the quaternion Fourier domain variable $\mathbf{u} = [u_1,u_2] \in \mathbb{R}^2$. The QFT becomes
\begin{equation}
   F(\mathbf{u}) = \int_{\mathbb{R}^2}e^{-\mathrm{i}2\pi u_1x_1}f(\mathbf{x})e^{-\mathrm{j}2\pi u_2 x_2}d^2\mathbf{x}.
\end{equation}
The corresponding $(N, N)$-point 2-D quaternion DFT (QDFT) of the discrete quaternion signal {$f(q,b)$} is 
\begin{equation}
    F_Q(r,s) =  {\frac{1}{N^2}} {\sum_{q=0}^{N-1}\sum_{b=0}^{N-1}}e^{-\mathrm{i}2\pi rq/N}f(q,b)e^{-\mathrm{j}2\pi sb/N}.
\end{equation}

Define the 2-D QDFT matrix $\mathbf{F}_Q \in \mathbb{H}^{n\times n}$ for a vectorized quaternion image $\mathbf{x}\in\mathbb{H}^{n}$, $n=N^2$, as $\mathbf{F}_Q = (\widetilde{\mathbf{F}}_Q^{(\mathrm{i})}\otimes  \widetilde{\mathbf{F}}_Q^{(\mathrm{j})} )\mathbf{x}$ where $\otimes$ denotes the Kronecker product. Here, $\widetilde{\mathbf{F}}_Q^{(\mathrm{\mu})}\in\mathbb{H}^{N\times N}$ denotes the N-point 1-D QDFT matrix for the \textit{quaternion unit} $\mathrm{\mu}$, whose entries are $\widetilde{\mathbf{F}}_Q^{(\mathrm{\mu})}(r,s) =\frac{1}{\sqrt{N}}e^{-\mu2\frac{\pi}{N} rs} $ and $\otimes $ denotes the Kronecker product. This matrix formulation of the 2-D QDFT is possible because of the separability of the two-sided QDFT. Computational implementations of the 2-D QDFT via quaternion fast FT (QFFT) are also available.

Figure \ref{fig:qfpr}c employs real data with an RGB image of size $N=128$. Here, $n=128^2=1024$ and $L=10$. 
An increase in the number of coding variables improves the reconstruction quality.  
Figure \ref{fig:qfpr}d shows a phase transition performance of QWF for the Fourier QPR. The results are averaged over $100$ Monte-Carlo simulations for recovering a random quaternion signal with $n=1000$. 

			\end{multicols}
			
			\begin{center}
				\includegraphics[width=\linewidth]{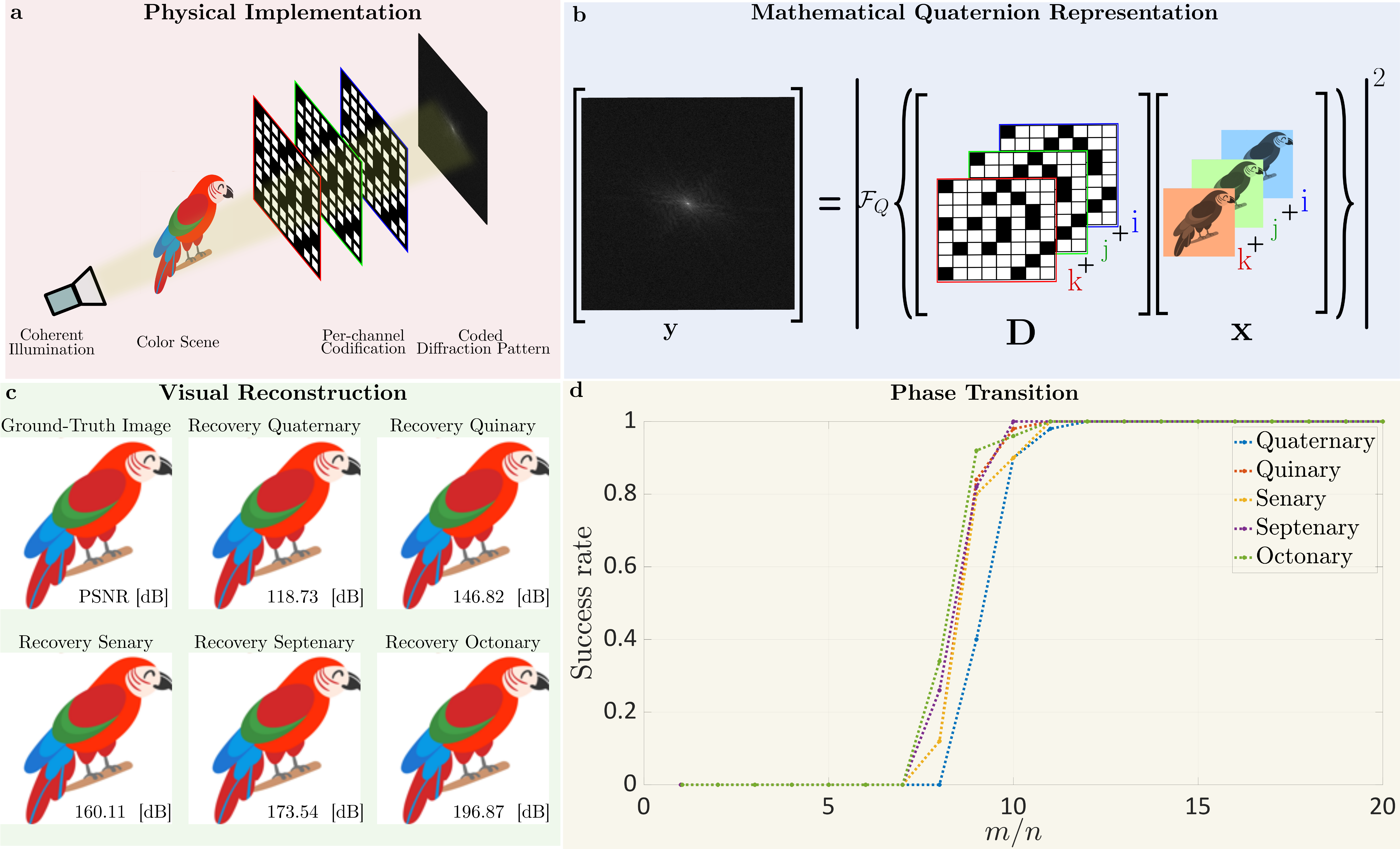}
				\captionof{figure}{(a) Illustration of the physical measurement set-up of Fourier QPR in optical imaging where a coherent illumination diffracts from an object. A DOE codifies the scene. Considering far-zone propagation, the sensor measures the magnitude of the coded scenes via QDFT. (b) Quaternion modeling of Fourier phase retrieval, where the sensing matrix contains both the per-channel codification of the DOE and the quaternion Fourier transform. (c) Fourier QPR recovery of an RGB image with a different number of coding elements from $d=4$ to $d=8$ using QWF. Here, $m/n=L=10$ and $n=128^2$. (d) Phase transition of QWF for Fourier QPR by varying the sample complexity $m/n$ and different values of $d$. Here, we set $n=1000$.}
				\label{fig:qfpr}
			\end{center} 
		\end{tcolorbox}
\end{strip}
X-ray crystallography. In diffractive optical imaging (DOI), the sensing process is modeled according to the propagation zone \cite{pinilla2023unfolding}. For instance, far-zone DOI comprises a DOE that modulates the input wavefront, and propagation is modeled by the Fourier transform of the coded wavefront (see Fourier QPR box, Figure \ref{fig:qfpr}a). 
While we illustrate only the far-zone case here, other zones (mid- and near-) are addressed similarly using a suitable sensing matrix.

Consider a two-dimensional (2-D) DOE $\mathbf{T}^k \in \mathbb{C}^{N\times N}$, where $k$ is the snapshot index $k=1,\dots,L$. 
Every entry of the DOE is usually a random variable on the set of possible coding elements $\mathcal{D}=\{c_1,\dots,c_d\}$ with probability $\frac{1}{d}$ for each variable and $d$ denotes the number of possible coding elements. In conventional complex-valued PR \cite{candes2015phase_2}, the set of coding variables are $\mathcal{D} = \{1,-1,-\mathrm{i},\mathrm{i}\}$. There are several recent works on the uniqueness guarantees of PR obtained by properly designing the DOE \cite{pinilla2023unfolding}. The sensing matrix here is $\mathbf{A} = \left[(\mathbf{D}^1)^H \mathbf{F}^H,\dots,(\mathbf{D}^L)^H \mathbf{F}^H\right]^H$, where $\mathbf{D}^k \in \mathbb{C}^{n\times n}$ is a diagonal matrix with the vectorized DOE $\mathbf{T}^k$ and $\mathbf{F}\in\mathbb{C}^{n\times n}$  is a 2-D DFT matrix.

 {We aim to develop the hypercomplex version of this problem, leading thus to new algorithmic opportunities by harnessing hypercomplex algebras. Thus, the Fourier QPR is formulated using one definition of the 2-D QDFT matrix $\mathbf{F}_Q$.} 
Denote the set of possible quaternion coding variables by  {$\mathcal{D}_Q = \{q_1,\dots,q_d\}$},  {where $q_i \in \mathbb{H}$ for $i=1,\dots,d$}. Then, the DOE of the $k$-th snapshot $\mathbf{T}^k_Q \in \mathbb{H}^{N\times N}$ is built upon the quaternion coding variables. Assume $\mathbf{D}^k_Q \in \mathbb{H}^{n\times n}$ to be the diagonalization of the quaternion DOE. Based on this formulation, the Fourier QPR measurements are
\begin{equation}
    \mathbf{y} = \left\vert \overbrace{\left[(\mathbf{D}^1_Q)^H\mathbf{F}_Q^H,\dots,(\mathbf{D}^L_Q)^H\mathbf{F}_Q^H\right]^H}^{\mathbf{A}\in \mathbb{H}^{Ln\times n}}\mathbf{x}\right\vert ^2 
\end{equation}
where $\mathbf{y} \in \mathbb{R}^{nL}$ are the phaseless measurements; see Figure \ref{fig:qfpr}b) for a visual representation of this formulation. 

 To recover the quaternion signal $\mathbf{x}$ from $\mathbf{y}$, we solve the optimization problem \begin{equation}
     \minimize_{\widetilde{x}\in\mathbb{H}^n} \frac{1}{2nL} \sum_{\ell=1}^{{nL}}(\vert\mathbf{a}_\ell^*\widetilde{\mathbf{x}}\vert^2-y_\ell)^2,
 \end{equation} where $\mathbf{a}_\ell = \mathbf{f}_{\alpha_\ell}^H(\mathbf{D}_Q^{k_\ell})^H$ is a row of the QDFT matrix with $\alpha_\ell = \lfloor \frac{\ell}{n}\rfloor+1$ and $k_\ell = ((\ell -1)\mod n) +1$. This formulation is also solved using QWF. In Figure \ref{fig:qfpr}c) we tested the algorithm with real data, an RGB image. Although all results produce satisfactory reconstructions, increasing the number of coding variables improves the reconstruction quality.  {Figure \ref{fig:qfpr}d) shows a phase transition performance of QWF, wherein we declare success when $d(\mathbf{x},\widetilde{\mathbf{x}})<1e-5$.} The results show that increasing the number of coding variables allows for better performance in QWF. This allows us to conclude that using quaternion approaches, increases the coding richness compared with the complex-valued Fourier PR which translates to better recovery performance. 
 
 The Fourier QPR may also be extended to octonions by following the octonion Fourier transform (OFT) \cite{blaszczyk2020discrete} defined for tri-variate functions. This enables novel 3-D PR applications, such as 3-D object reconstruction, structured light 3-D object sensing, and digital holography.  {Similar to QFT, OFT has no unique definition. Here, we follow the definition in \cite{blaszczyk2017octonion} because of its several useful properties (similar to complex-valued FT) such as Hermitian symmetry, shift theorem, Plancherel-Parseval theorem, and derivative theorem.} Consider the octonion continuous-valued signal $f(\mathbf{x}): \mathbb{R}^3\rightarrow \mathbb{H}, \mathbf{x} = [x_1,x_2,x_3] \in \mathbb{R}^3$, whose OFT $F(\mathbf{u})=\mathcal{F}_O\{f(\mathbf{x})\}$, where ${F}(\cdot)$ is a function of the octonion Fourier domain variable $\mathbf{u} = [u_1,u_2,u_3] \in \mathbb{R}^3$. The OFT is given by
\begin{align}
&\mathcal{F}_O\{(f(\mathbf{x})\}=\nonumber\\ & F(\mathbf{v}) = \int_{\mathbb{R}^3} f(\mathbf{x})e^{-\mathrm{e}_1 2\pi f_1 x_1}e^{-\mathrm{e}_2 2\pi f_2 x_2}e^{-\mathrm{e}_4 2\pi f_3 x_3}d^3\mathbf{x}.
\end{align}
 
 Here, only the imaginary units $e_1,e_2,e_4$ are used because, following the power expansion of the exponential function, this product yields a full octonion function. Define the octonion DFT (ODFT) for a discrete octonion signal $f(\mathbf{n})=f(n_1,n_2,n_3)$ with  {$n_1,n_2,n_3 \in \{0,\dots, N-1\}$ as
\begin{align}
     &F_O(k_1,k_2,k_3) =\frac{1}{N^3}\sum_{n_1=0}^{N-1}\sum_{n_2=0}^{N-1}\sum_{n_3=0}^{N-1}f(n_1,n_2,n_3)\times \nonumber\\&e^{-\mathrm{e}_1 2\pi k_1 n_1/N}e^{-\mathrm{e}_2 2\pi k_2 n_2/N}e^{-\mathrm{e}_4 2\pi k_3 n_3/N}.
 \end{align}}
 Similar to Fourier QPR, we have the corresponding ODFT sensing matrix $\mathbf{F}_O \in \mathbb{O} ^{n \times n}$. The DOE matrix $\mathbf{T}_O^k\in\mathbb{O}^{N\times N}$ has more coding variables. The design of optimal DOE to guarantee unique recovery in Fourier HPR is currently an unaddressed problem.

\subsection{Hypercomplex STFT PR}
The STFT of a signal provides the spectral characteristics localized in time. It is obtained by partitioning the signal into overlapping segments and then taking the windowed Fourier transform of each segment. Conventional PR has shown that, for the same number of measurements, the STFT magnitude measurements lead to better PR performance than an over-sampled DFT. The STFT PR applications include ultrashort laser pulses or Fourier ptychography. Traditional complex-valued STFT PR problem considers a discrete-time signal $\mathbf{x} = [x_1,\dots, x_n]^T \in \mathbb{C}^{n}$ and window $\mathbf{w} = [w_1,\dots,w_T]^T\in\mathbb{C}^{T}$ such that the magnitude-only measurements using the discrete-time STFT of the $\mathbf{x}$ respect to the window $\mathbf{w}$ are \begin{equation}
    \mathbf{Y}_{r,s} = \left\vert\sum_{q=1}^{N} \mathbf{x}[q] \mathbf{w}[rL-q]e^{-\mathrm{i}2\pi sq/N}\right\vert^2,\end{equation} $s = 1,\dots,N$, $r = 1,\dots,R$,  where $R = \left \lceil{\frac{N+T-1}{L}
}\right \rceil$ is the number of short-time sections and  {$\mathbf{w}[n]$ is periodically extended over the boundaries of the window}. The choice of the window affects the uniqueness guarantees of recovery. 

In the case of STFT QPR, the continuous-valued quaternion STFT (QSTFT) of a quaternion  {2-D} signal $f(\mathbf{x}) \in  \mathbb{H}$ respect to the window function $w(\mathbf{x}) \in \mathbb{H}$ is 
\begin{align}&F_{QSTFT}(\mathbf{v},\mathbf{u}) \nonumber\\&= \int_{\mathbb{R}^2} f(\mathbf{x}) w(\mathbf{x}-\mathbf{u}) e^{-\mathrm{i}2\pi x_1v_1} e^{-\mathrm{j}2\pi x_2v_2}d^2\mathbf{x},\end{align}
where $\mathbf{v}$ and $\mathbf{u}$ are variables in the STFT domain. Its corresponding discrete version with a quaternion window $w(p,q) \in \mathbb{H}$  and a discrete signal $f(p,q)\in \mathbb{H}$, $p,q \in {1,\dots,N}$, is 
\begin{align}&F(\mathbf{v},\mathbf{u}) \nonumber\\&=  {\frac{1}{N^2}\sum_{q=0}^{N-1}\sum_{b=0}^{N-1}}f(q,b)w(p-u_1,q-u_2)e^{-\mathrm{i}2\pi qr/N}e^{-\mathrm{j}2\pi sb/N}.\end{align}
In a matrix representation, the QSTFT measurements are $\mathbf{Y}_{r,s} =  \mathbf{f}_s^*\mathbf{W}_r\mathbf{x}$ where $\mathbf{W}_r\in \mathbb{H}^{N\times N}$ is a diagonal matrix created by shifting the quaternion vector $\mathbf{w}_r\in \mathbb{H}^T$ across the diagonal and $\mathbf{f}_s^*$ is the $s$-th row of the 2-D QDFT matrix. The QSTFT PR problem entails recovering $\mathbf{x}$ by solving the optimization problem: 
\begin{equation}\minimize_{\widetilde{\mathbf{x}}\in \mathbb{H}^N} \frac{1}{2nL} \sum_{r=1}^{R}\sum_{s=1}^{N}(\vert\mathbf{f}_s^*\mathbf{W}_r\widetilde{\mathbf{x}}\vert^2-\mathbf{Y}_{r,s})^2.\end{equation}

The STFT OPR is similarly defined using the octonion STFT (OSTFT) for 3-D signals as 

\begin{align}
    &F_{OSTFT}(\mathbf{v},\mathbf{u})=\nonumber\\&\int_{\mathbb{R}^3}f(\mathbf{x})w(\mathbf{x}-\mathbf{u}) e^{-\mathrm{e}_1 2\pi x_1v_1} e^{-\mathrm{e}_22\pi x_2v_2}e^{-\mathrm{e}_42\pi x_3v_3}d^3\mathbf{x}.
\end{align}

\subsection{Hypercomplex Wavelet PR}
Wavelet PR has shown promising applications in audio processing, and its hypercomplex variant may be formulated via the quaternion and octonion wavelet transforms. The quaternion wavelet transform (QWT) of a quaternion-valued signal $f(\mathbf{x}):\mathbb{R}^2\rightarrow \mathbb{H}$ with respect to the mother \textit{quaternion wavelet} $\psi(\mathbf{x}):\mathbb{R}^2\rightarrow\mathbb{H}$ for a scaling factor $a$, a rotation angle $\theta$, and the coordinates in the wavelet domain $\mathbf{b} \in \mathbb{R}^2$ is 
\begin{equation}
F_{QWT} (\psi,a,\theta,\mathbf{b}) = \int_{\mathbb{R}^2} f(\mathbf{x})\frac{1}{a}\psi\left( \mathbf{R}_{-\theta} \frac{1}{a} \left(\mathbf{x-b}\right) \right)d\mathbf{x}^2,\end{equation}
where $\mathbf{R}_\theta \in \mathbb{R}^{2\times 2}$ is the rotation matrix on $\mathbb{R}^2$ defined as $\mathbf{R}_\theta = \left(\begin{smallmatrix}
    \cos(\theta) & \sin(\theta)\\-\sin(\theta) & \cos(\theta)
    \end{smallmatrix}\right)$. We can interpret this definition as the convolution of the scaled and rotated mother quaternion wavelet $\psi(\mathbf{x})$ with the 2-D quaternion signal ${f}(\mathbf{x})$. It is instructive to define the family of wavelets $\{\psi^k(\mathbf{x})\}_{k=1}^L$, where $\phi^k(\mathbf{x}) = \frac{1}{a^k}\psi\left( \mathbf{R}_{-\theta^k}\frac{\mathbf{x}}{a^k}\right) $ and $L$ is the number of transformations of the mother wavelet. Then, the QWT becomes $F_{QWT}(\psi,\mathbf{b}) = \{f(\mathbf{x})\star\phi_j(\mathbf{b})\}_{k=1}^{L}$ where $\star$ is the convolution operation. The discrete QWT (DQWT) considers the sampled versions of the signal as of $f(p,q)$ and the mother wavelet transformation $\psi^k(p,q)$, $p,q =1,\dots, N$, as 
    \begin{equation}
         F(r,s) = \sum_{p=1}^{N}\sum_{q=1}^{N} f(p,q)\psi^k(r-p,s-q),
    \end{equation}
    for $r,s=1,\dots,N$. 
    
In the hypercomplex wavelet QPR problem, the magnitude-only measurement vector becomes $\mathbf{y} = \left\vert\left[(\bsym{\psi}^1\star\mathbf{f})^T,\dots,(\bsym{\psi}^L\star\mathbf{f})^T\right]^T \right\vert^2$, where $\mathbf{f}\in\mathbb{H}^n$ is the vectorization of the quaternion signal and $\bsym \psi^k \in \mathbb{H}^n$ the vectorization of the $k$-the wavelet. The HPR then entails solving the optimization problem 
\begin{equation}
    \minimize_{\widetilde{\mathbf{x}}\in\mathbb{H}^n} \frac{1}{2nL}\sum_{k=1}^{L} \left(\left\vert\bsym{\psi}^k\star\widetilde{\mathbf{x}} \right\vert^2 - \left\vert\bsym{\psi}^k\star\mathbf{f} \right\vert^2\right)^2.
\end{equation}
The selection of the quaternion mother wavelet may reduce the ill-posedness of this QWT PR. For instance, quaternion Haar functions have shown promising applications in multi-resolution image analysis. Similarly, quaternion Gabor wavelets arise in applications such as optical flow estimation. Some common choices of quaternion mother wavelets are built upon combining the QFT and conventional complex-valued wavelets such as Haar, biorthogonal, Daubechies, and Coiflet. Similar problems are found in octonion wavelet PR that employs octonion WT. 

\section{Summary and Future Outlook}
We presented an emerging framework of HPR for high-dimensional signals with several applications in optical imaging. Conventional PR algorithms and formulations are derived for their corresponding hypercomplex versions. By harnessing the Clifford algebra of hypercomplex numbers, HPR yields better performance and novel theoretical guarantees compared with the conventional complex-valued PR methods.

Corresponding to several variants of complex-valued PR, a large number of HPR problems may be formulated using, for example, a hypercomplex LCT matrix.  {We formulated these problems for specific definitions of QFT and OFT. Analyzing other QFT and OFT definitions remains an open problem.} Moreover, some PR problems require bispectrum analysis where the underlying signal is non-Gaussian and nonlinear as in radar, microscopy, and holography. The bispectrum HPR arises in color object recognition. Finally, Fractional Fourier transform (FrFT)-based PR has found applications in optical imaging and radar \cite{pinilla2021wavemax}, wherein the challenge is to resolve the additional rotation ambiguity FrFT PR. The HPR variant may be formulated using the quaternion FrFT. 

Further, methods that promote a given structure on the signal such as low-rank or sparsity may be applied to HPR to improve the recovery performance. These approaches follow from the use of hypercomplex applications to popular inverse problems such as low-rank matrix representation of RGB image denoising or sparse representation in MSI reconstruction. Following the recent advances in data-driven phase retrieval methods, there are also exciting opportunities in learning-based PR using quaternion and octonion deep learning networks. Finally, the co-design of diffractive optical elements and hypercomplex reconstruction algorithms to enable more precise reconstructions remains an open HPR problem.

\bibliographystyle{IEEEtran}
\bibliography{references}

\end{document}